\documentclass[preprint,onecolumn,aps,nofootinbib,notitlepage,bobalancelastpage]{revtex4-2}
\usepackage{amsmath}
\usepackage{amsfonts}
\usepackage{amssymb}
\usepackage{mathtools}
\usepackage{graphicx}
\usepackage{xcolor}
\usepackage[colorlinks=True,citecolor=black,linkcolor=black,urlcolor=black]{hyperref}
\usepackage{hypcap}
\usepackage{braket}
\usepackage{yfonts}
\usepackage{bm}
\usepackage[mathscr]{eucal}
\newcommand\eq[1]{\begin{align}#1\end{align}}

\newcommand\nh{{N_\mathcal{H}}}

\definecolor{myBlue}{RGB}{31,119,180}
\definecolor{myOrange}{RGB}{255,127,14}
\definecolor{myGreen}{RGB}{44,160,44}
\definecolor{myRed}{RGB}{214,39,40}
\definecolor{myPurple}{RGB}{148,103,189}

\makeatletter
\def\p@figure{\color{myBlue}}
\def\p@equation{\color{myRed}}
\makeatother

\begin{document}

\title[Anatomy of localisation protected quantum order on Hilbert space]{Anatomy of localisation protected quantum order on Hilbert space}

\author{Sthitadhi Roy}
\email{sthitadhi.roy@icts.res.in}
\affiliation{International Centre for Theoretical Sciences, Tata Institute of Fundamental Research, Bengaluru 560089, India}

\begin{abstract}
Many-body localised phases of disordered, interacting quantum systems allow for exotic localisation protected quantum order in eigenstates at arbitrarily high energy densities. 
In this work, we analyse the manifestation of such order on the Hilbert-space anatomy of eigenstates. 
Quantified in terms of non-local Hilbert-spatial correlations of eigenstate amplitudes, we find that the spread of the eigenstates on the Hilbert-space graph is directly related to the order parameters which characterise the localisation protected order, and hence these correlations, in turn, characterise the order or lack thereof. Higher-point eigenstate correlations also characterise the different entanglement structures in the many-body localised phases, with and without order, as well as in the ergodic phase. The results pave the way for characterising the transitions between many-body localised phases and the ergodic phase in terms of scaling of emergent correlation lengthscales on the Hilbert-space graph.
\end{abstract}

\maketitle

\newpage

\tableofcontents

\newpage

\section{Introduction}
The many-body localised (MBL) phase of disordered quantum systems is a fundamentally new dynamical phase of quantum matter wherein an isolated quantum system fails to thermalise under dynamics (see Refs.~\cite{nandkishore2015many,abanin2017recent,alet2018many,abanin2019colloquium} for reviews and references therein). These systems violate the eigenstate thermalisation hypothesis (ETH)~\cite{deutsch1991quantum,srednicki1994chaos,rigol2008thermalisation,dalessio2016from,deutsch2018eigenstate} and hence, fall outside the conventional paradigm of equilibrium statistical mechanics and thermodynamics. 

One of the most important, and interesting, consequences of this is that the interplay of symmetries and localisation can lead to the presence of exotic quantum order at arbitrary energy densities, which is forbidden in ergodic systems due to ETH~\cite{huse2013localisation,pekker2014hilbert,kjall2014many,prakash2017eigenstate,friedman2018localisation,roy2018dynamical, roy2018nonequilibrium,paramesawran2018many,sahay2021emergent}. This kind of order is often eponymously dubbed as {\it eigenstate order} as the order is a property of the highly excited eigenstates themselves and not the usually considered thermodynamic equilibrium state. Localisation is key to such order as it ensures that these highly excited eigenstates have area-law entanglement~\cite{serbyn2013local,bauer2013area} and thus have the possibility of hosting order usually observed in gapped ground states of quantum systems with local Hamiltonians~\cite{hastings2007area}.
In fact, these ideas can be extended to manifestly out-of-equilibrium settings such as periodically-driven systems where order, such as time-crystalline, can be realised which have no analogue in time-independent Hamiltonian systems~\cite{khemani2016phase,else2016floquet,yao2017discrete,moessner2017equilibration}. As the MBL systems are characterised by an extensive set of quasi-local integrals of motion~\cite{serbyn2013local,huse2014phenomenology,ros2015integrals} and the MBL protected order, if present, is characterised by local order parameters, much of the studies on such MBL systems, both with and without order, is based on the spectral and dynamical properties of real-space local observables.

\begin{figure}[!b]
\includegraphics[width=0.7\linewidth]{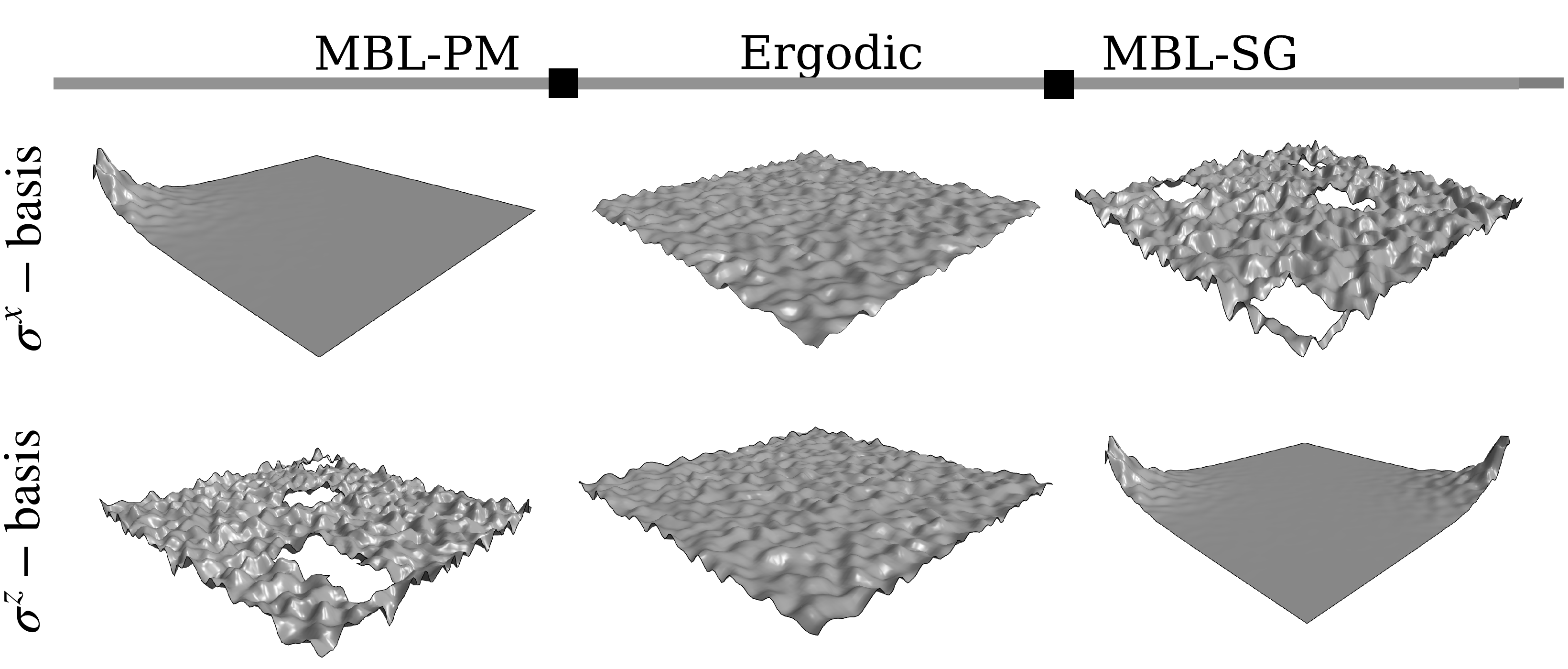}
\caption{Schematic, visual summary of the Hilbert-space anatomy of the eigenstates in the three phases of the model in Eq.~\ref{eq:ham}, MBL-paramagnet (MBL-PM), ergodic, and MBL-spin glass (MBL-SG) in the two bases, $\sigma^x$- and $\sigma^z$- product state bases (top and bottom rows respectively). In the MBL-PM phase, the eigenstates are conventionally localised in the the $\sigma^x$-basis manifested in a single peak in the amplitudes. In the MBL-SG phase, the system breaks a $\mathbb{Z}_2$ symmetry such that eigenstates are cat states of approximate $\sigma^z$-product states manifested in the bilocalised structure in the $\sigma^z$-basis. The MBL-PM and MBL-SG eigenstates in the $\sigma^z$- and $\sigma^x$-basis on the other hand, are spatially spread out over the entire Hilbert space but with holes in them such that they are weakly multifractal. In the ergodic phase, the eigenstates are fully extended in either basis.}
\label{fig:summary}
\end{figure}

A parallel approach to studying MBL systems which has proven extremely useful in the recent past has been to map the problem onto a disordered hopping problem on the complex, correlated Hilbert/Fock-space graph of the system~\cite{altshuler1997quasiparticle,basko2006metal,MonthusGarel2010PRB,deluca2013ergodicity,serbyn2015criterion,pietracaprina2016forward,baldwin2016manybody,logan2019many,roy2018exact,roy2018percolation,roy2019self,mace2019multifractal,pietracaprina2021hilbert,detomasi2019dynamics,ghosh2019manybody,nag2019manybody,roy2020fock,biroli2020anomalous,tarzia2020manybody,detomasi2020rare,hopjan2020manybody,tikhonov2021eigenstate,roy2021fockspace,roy2020strong,sutradhar2022scaling,roy2022hilbert,sutradhar2022scaling,roy2023diagnostics,creed2023probability}.
This approach, in fact, is quite natural as the eigenstates fundamentally live on the many-body Hilbert space and dynamical correlations or propagators therein naturally carry a lot of information. Some of the key insights that have emerged from this approach are multifractality of MBL eigenstates~\cite{deluca2013ergodicity,luitz2015many,mace2019multifractal,roy2021fockspace}, Kosterlitz-Thouless-like scaling near the many-body localisation transition~\cite{mace2019multifractal,roy2021fockspace} consistent with phenomenological theories~\cite{goremykina2019analytically,dumitrescu2018kosterlitz,morningstar2019renormalization,morningstar2020manybody}, strong correlations in the effective Hilbert/Fock-space disorder as a necessary condition for an MBL phase to exist~\cite{roy2020fock,roy2020localisation}, and connections between effective Hilbert/Fock-space lengthscales and real-space local observables~\cite{detomasi2020rare,roy2021fockspace}. However, the Hilbert-space anatomy of MBL eigenstates with localisation protected order is hitherto unexplored and constitutes the central interest of this work. Precisely stated, we ask the question how does the anatomy of eigenstates on the Hilbert space,  quantified via Hibert-spatial correlations of eigenstate amplitudes, distinguish between MBL phases with and without localisation protected order as well as distinguish them from an ergodic phase.

Our analysis shows that the anatomy of eigenstates across the three phases is sharply different, see Fig.~\ref{fig:summary} for visual summary. More importantly, the spatial spread of the eigenstates on the Hilbert-space graph can be directly related to the order parameters such as the Edwards-Anderson spin-glass order parameter and the spin polarisations in the eigenstates. This results in a sharp characterisation of the eigenstate phases in terms of Hilbert-space correlations. We also find that the emergent correlation lengths on the Hilbert-space graph are directly related to the order parameters and indeed, the divergence of these correlation lengths is consonant with the vanishing of these order parameters. This suggests that a scaling theory in terms of these correlation lengths might shed light on the nature of the order-parameter scaling at the phase transitions. Since the transitions between the MBL phases and the ergodic phase is also an entanglement transition as the eigenstates transition from a volume-law bipartite entanglement entropy in the ergodic phase to an area-law in the MBL phases~\cite{bauer2013area}, we also find its manifestations in higher-point Hilbert-space correlations.

The rest of the paper is organised as follows. In Sec.~\ref{sec:model} we describe a model of a $\mathbb{Z}_2$ symmetric Ising spin chain and its phase diagram. In this section, we also describe the underlying Hilbert-space structure of the model. In Sec.~\ref{sec:hscorrorder}, we discuss the spatial correlations of eigenstate amplitudes on the Hilbert-space graph and show how they are directly related to the Edwards-Anderson spin-glass order parameter. The connection between the entanglement properties in the three phases and higher-point Hilbert-space correlations is elucidated in Sec.~\ref{sec:hscorrent}. We close with concluding remarks in Sec.~\ref{sec:conclusions}.

\section{Model and Hilbert-space structure \label{sec:model}}
\subsection{$\mathbb{Z}_2$ symmetric Ising spin chain}
As a concrete model exhibiting different kinds of localised phases, with and without localisation protected order, we consider a one-dimensional quantum spin-1/2 chain described by the Hamiltonian~\cite{sahay2021emergent}
\eq{
H = \sum_{i=1}^L [J_i\sigma^z_i\sigma^z_{i+1} + h_i\sigma^x_i + V_i(\sigma^x_i\sigma^x_{i+1}+\sigma^z_i\sigma^z_{i+2})]\,,
\label{eq:ham}
}
where $\sigma_i$ denotes the Pauli matrix on site $i$ and we use periodic boundary conditions by identifying site $L+i$ with $i$. The disordered exchange couplings and the fields are drawn randomly from uniform distributions, $J_i\in[-W_J,W_J]$, $h_i\in[-W_h,W_h]$, and $V_i\in[-W_V,W_V]$. We consider a parametrisation where $\sqrt{W_JW_h}=1$ is held fixed and vary the ratio $W_h/W_J$ to tune across different phases. The most important feature to note in the Hamiltonian in Eq.~\ref{eq:ham} is that it possesses a global $\mathbb{Z}_2$ symmetry effected by the parity operator $P = \prod_{i=1}^L\sigma^x_i$. The two distinct localised phases in the model are distinguished by order emerging out of spontaneous breaking of this symmetry or lackthereof. 

In the limit of $W_J\gg W_h,W_V$, the disordered Ising interactions, $\sigma^z_i\sigma^z_{i+1}$, dominate such that the eigenstates are weakly dressed versions of cat states of $\sigma^z$-product states. The physical states in a thermodynamically large system break the associated $\mathbb{Z}_2$ symmetry, and the system is said to be in the so-called MBL spin-glass (MBL-SG) phase. In finite-sized systems, the long-ranged spin-glass order is best captured via an Edwards-Anderson like order parameter~\cite{huse2013localisation,pekker2014hilbert,kjall2014many,sahay2021emergent} 
\eq{\chi = L^{-1}\sum_{i,j}\braket{\sigma^z_i\sigma^z_j}^2\,,\label{eq:EA}} where the expectation value is over eigenstates. This is different from the MBL phase in a model without the $\mathbb{Z}_2$ symmetry as the $z$-polarisation, $m_z = L^{-1}\sum_i\braket{\sigma^z_i}^2$, is vanishingly small owing to the `feline' nature of the eigenstates. On the other hand, for $W_h\gg W_J,W_V$, the disordered onsite fields which couple to the $\sigma^x$ operators dominate, and the model is in a MBL paramagnet (MBL-PM) phase which does not break the $\mathbb{Z}_2$ symmetry. In this phase the eigenstates are weakly dressed versions of $\sigma^x$-product states and are characterised by a finite $x$-polarisation, $m_x=L^{-1}\sum_i\braket{\sigma^x_i}^2$.

Note that for $W_V=0$, the model in Eq.~\ref{eq:ham}, via a Jordan-Wigner transformation, is equivalent to that of free fermions with disordered hoppings and disordered fields. It is therefore Anderson localised~\cite{anderson1958absence,mott1961theory} for any $W_J,W_h\neq 0$ and there is a direct transition at $W_J=W_h$ from the MBL-PM at $W_J/W_h<1$ to the MBL-SG at $W_J/W_h>1$. A finite $W_V$ however, makes the model non-integrable and more importantly, opens up an ergodic phase between the two MBL phases. As a result upon increasing $W_J/W_h$, the model first transitions from the MBL-PM to an ergodic phase and then to the MBL-SG phase; see Ref.~\cite{sahay2021emergent} for a detailed numerical study of the phase diagram. In this entire work we use $W_V=0.7$ for which the MBL-PM to ergodic transition happens at $(W_J,W_h) \approx (0.32,3.2)$ and the ergodic to MBL-SG transition happens at $(W_J,W_h) \approx (3.2,0.32)$. All numerical results shown in the rest of what follows is obtained from extracting a few, $\approx 50-100$, eigenstates from the middle of the spectrum of the Hamiltonian in Eq.~\ref{eq:ham} and averaged over $\approx 1000$ disorder realisations.
Having summarised the phases of the model, we next turn to the Hilbert-space structure.

\subsection{Hilbert-space structure of eigenstates}
The structure of the two localised phases hosted by the model furnishes two natural choices of (local) bases for the Hilbert space, namely spin configurations polarised in the $z$ direction and polarised in the $x$ direction. We will denote these states as $\ket{\vec{s}^z_\alpha}$ and $\ket{\vec{s}^x_\alpha}$ respectively where $\sigma^\mu_i\ket{\vec{s}^{\,\mu}_\alpha}=s^\mu_{i,\alpha}\ket{\vec{s}^{\,\mu}_\alpha}$ with $s^\mu_{i,\alpha}=\pm 1$, and $\alpha=1,2,\cdots,\nh$ where $\nh=2^L$ is the Hilbert-space dimension. Both these basis choices endow the respective Hilbert-space graphs with a distance which is the number of spins anti-aligned (in their respective directions) between two states. Formally, these distances, denoted by $r_{\alpha\beta}^{\mu}$ are given by
\eq{
  r_{\alpha\beta}^{\mu} = \frac{1}{4}\sum_{i=1}^L (s^\mu_{i,\alpha}-s^\mu_{i,\beta})^2 = \frac{L}{2}-\frac{1}{2}\sum_{i=1}^L s^\mu_{i,\alpha}s^\mu_{i,\beta}\,.
  \label{eq:dist}
}
This notion of distance also implies that the number of states $\ket{\vec{s}^{\,\mu}_\beta}$ that are at distance $r$ from any given state $\ket{\vec{s}^{\,\mu}_\alpha}$ is given by $N_r\equiv \binom{L}{r}$.

In order to understand the structure of the phases on the Hilbert-space graph, one of the most fundamental quantities is the non-local propagator
\eq{
  G_{\alpha\beta}^\mu(\omega)= \braket{\vec{s}^{\,\mu}_\alpha|[\omega+i0^+-H]^{-1}|\vec{s}^{\,\mu}_\beta}\,.
  \label{eq:Gab}
}
The non-local propagator quantifies how the quantum states are spread out on the Hilbert-space graph and also encodes how the amplitudes are correlated with each other. As such, it is the basic building block for the dynamical response of any quantum state.
In fact, the residue of the non-local propagator in Eq.~\ref{eq:Gab} at an eigenenergy $E$ is intimately connected to a correlation function between the Hilbert-space amplitudes of the corresponding eigenstate $\ket{\psi}$ of the Hamiltonian. Decomposing the eigenstate in either basis as
\eq{
  \ket{\psi} = \sum_{\alpha=1}^{\nh}\psi_\alpha^\mu\ket{\vec{s}^{\,\mu}_\alpha}\,,
}
the relation between the Hilbert-space correlations and the non-local propagator can be explicitly expressed as 
\eq{
  |\mathrm{Res}_E G^\mu_{\alpha\beta}(\omega)|^2 = |\psi_\alpha^\mu|^2|\psi_\beta^\mu|^2\,.
}
This in turn can be recast as a spatial correlation on the Hilbert-space graphs by using the notion of distance on the Hilbert-space graph, defined in Eq.~\ref{eq:dist}, as
\eq{
  F^\mu(r) = \sum_{\substack{\alpha,\beta:\\ r^\mu_{\alpha\beta}=r}}|\psi_\alpha^\mu|^2|\psi_\beta^\mu|^2\,.
  \label{eq:Fr}
}
It is useful to note here that $F^\mu(r)$ can be interpreted as a bonafide probability distribution over $r$ as $F^\mu(r)\ge 0$ for all $r$ and $\sum_r F^\mu(r) = 1$.

\begin{figure}[!t]
\includegraphics[width=0.7\linewidth]{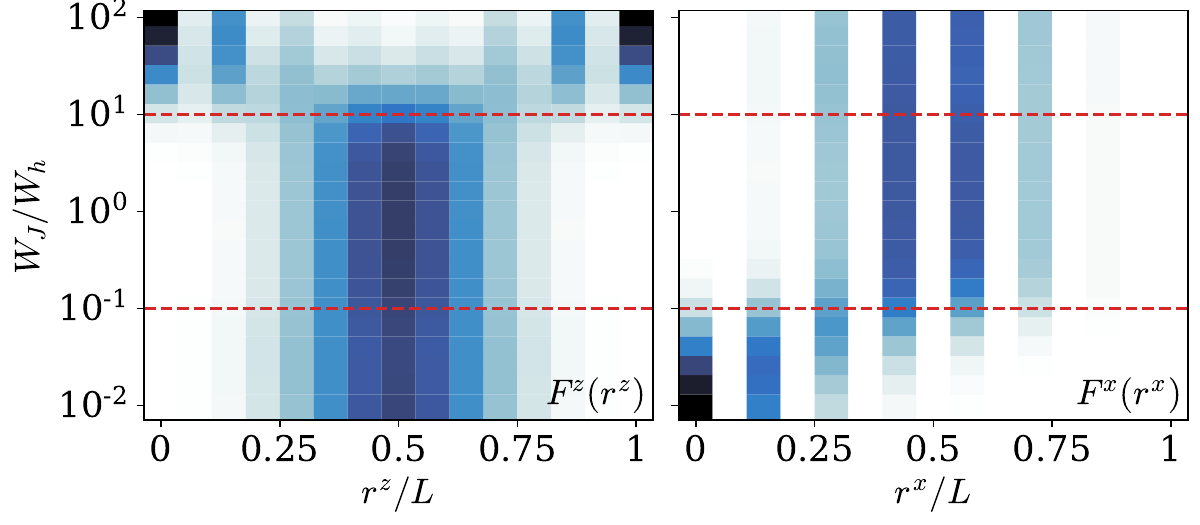}
\caption{The Hilbert-spatial correlation function $F^z(r)$ (left) and $F^x(r)$ (right), defined in Eq.~\ref{eq:Fr}, presented as a heatmap as a function of $r/L$ and $W_J/W_h$ for $L=14$. At the ergodic to MBL-SG phase transition at $W_J/W_h\approx 10$, the unimodal structure of $F^z(r)$ sharply changes to a bimodal structure indicating the bilocalised structure of the states consistent with their cat-like nature. At the MBL-PM to ergodic transition at $W_J/W_h\approx 0.1$, the peak in $F^x(r)$ sharply shifts from value $r/L <1/2$ to $r/L=1/2$. Note that $F^x(r)$ is identically zero for all odd $r^x$ as states separated by odd $r^x$ belong to different parity sectors.}
\label{fig:Fr}
\end{figure}

Before delving into their quantitative analysis, we present in Fig.~\ref{fig:Fr}, a broadbrush view of the correlation functions $F^z(r^z)$ and $F^x(r^x)$ by plotting them as heatmaps in the plane of $r/L$ and $W_J/W_h$. 
In the ergodic phase, the eigenstates are uniformly delocalised over the entire Hilbert space in either basis; as such $F^\mu(r)$ essentially mirrors the binomial profile of $N_r$ with a peak at $r/L = 1/2$.
In the MBL-SG phase, the cat-like nature of the eigenstates in the $\sigma^z$-basis is manifested in a bimodal structure of $F^z(r^z)$ with the peaks at $r^z=0$ and $r^z=L$ signifying that the states are bilocalised on $\sigma^z$-configurations which are related to each other by the parity operator. In the MBL-PM phase, $F^x(r^x)$ is peaked around a value of $r/L<1/2$. This is essentially an outcome of the fact that the eigenstates are exponentially localised around a $\sigma^x$-configuration. However, the number of states at distance $r$ from the localisation centre grows as $\binom{L}{r}$ (for $r\leq L/2$). As such, the combination of the exponential decay of the eigenstate amplitudes and the growth of number sites leads to $F^x(r^x)$ being peaked at a $r^x=pL/2$ where $p<1$~\cite{roy2021fockspace}.
From the numerical results in Fig.~\ref{fig:Fr}, it appears that $F^z$ in the MBL-PM phase and $F^x$ in the MBL-SG are very similar to that in the ergodic phase. As we show shortly, while this is indeed true as far as the overall Hilbert-spatial spread of the eigenstates is concerned, there are some subtle differences in their finer structure.
 
One of the quantities which is sensistive to these differences is $F^\mu(r=0)$, which is simply the inverse participation ratio (IPR) of the eigenstate in the $\sigma^\mu$-basis,
\eq{
  \mathcal{I}^\mu=\sum_{\alpha}|\psi_{\alpha}^\mu|^4\sim \nh^{-\tau^\mu}\,.
  \label{eq:IPR}
}
The scaling of $\mathcal{I}^\mu$ with $\nh$ via the exponent $\tau^\mu$ is a rather useful way of characterising an MBL phase and distinguishing it from an ergodic phase. The ergodic phase is characterised by extended eigenstates fully delocalised over the entire Hilbert-space such that $\tau^\mu=1$. On the other hand, in conventional models with only one kind of MBL phase (where, say, the disorder couples to the $\sigma^z$-operators) the MBL phase is characterised by (multi)fractal eigenstates on the Hilbert-space with $0<\tau^z<1$~\cite{deluca2013ergodicity,luitz2015many,mace2019multifractal,roy2021fockspace}.
\begin{figure}[!b]
\includegraphics[width=0.7\linewidth]{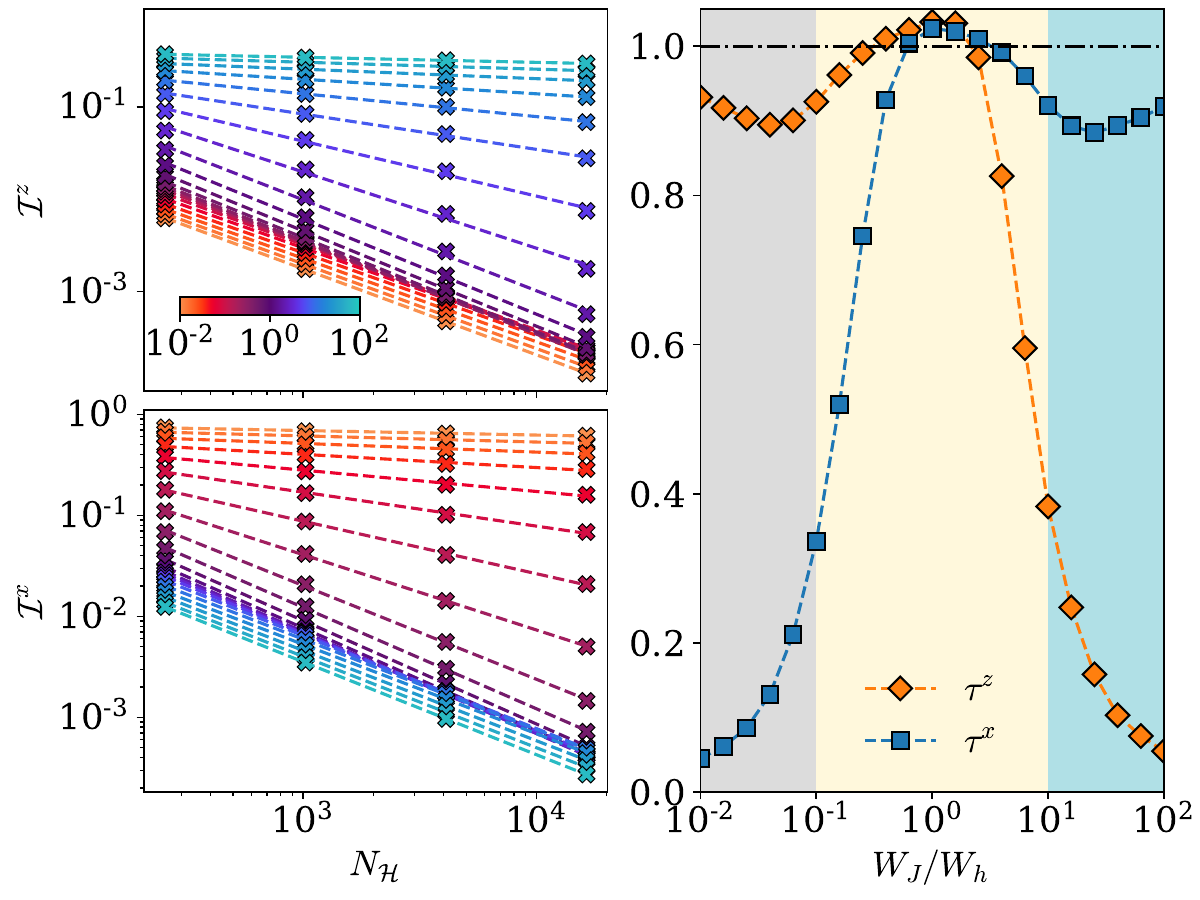}
\caption{Scaling of the IPRs defined in Eq.~\ref{eq:IPR} with the Hilbert-space dimension for different values of $W_J/W_h$. The upper left and lower left panels show $\mathcal{I}^z$ and $\mathcal{I}^x$ as function of $\nh$ where different colours correspond to different values of $W_J/W_h$ as indicated by the colourscale. The dashed lines are straight line fits on the double logarithmic scale, the slopes of which are used to extract the exponents $\tau^z$ and $\tau^x$. The exponents so extracted are shown in the right panel.}
\label{fig:ipr}
\end{figure}

For a model with different kinds of MBL phases, such as the one considered here, one has to be more careful and analyse the IPR in both the bases. The numerical results for the IPRs are shown in Fig.~\ref{fig:ipr}. In the ergodic phase, $0.1\lesssim W_J/W_h\lesssim 10$, the eigenstates can be well approximated as Gaussian random states such that both $\tau^z\approx 1$ and $\tau^x\approx 1$; this is indeed consistent with the numerical results. In the MBL-PM, since the eigenstates are weakly dressed versions of the $\sigma^x$-product states, one expects $\mathcal{I}^x$ to show clear fractal scaling with $\tau^x<1$ which the numerical results do confirm. In the same spirit, the eigenstates in the MBL-SG phase are weakly dressed versions of states bilocalised on two $\sigma^z$-configurations related to each other by the parity operator. One thus again expects $0<\tau^z<1$ in the MBL-SG phase.

The more suprising result, perhaps, is that the eigenstates in the MBL-PM phase are weakly fractal in the $\sigma^z$-basis and similarly, the eigenstates in the MBL-PM phase are weakly fractal in the $\sigma^x$-basis while the the results in Fig.~\ref{fig:Fr} might have suggested that they are fully delocalised. This fractality can be understood via the fact that the tranformation between the $\sigma^z$- and $\sigma^x$-bases can be effected by the Walsh-Hadamard transform. The key point is that the Walsh-Hadamard transform can be viewed as a Fourier transform on the Boolean group $(\mathbb{Z}/2\mathbb{Z})^L$. Formally this implies
\eq{
  \psi_\alpha^x = \frac{1}{\sqrt{\nh}}\sum_\beta \psi_\beta^z \, (-1)^{\vec{s}^{\,x}_\alpha \odot \vec{s}^{\,z}_\beta}\,,
  \label{eq:wht}
}
where ${\vec{s}^{\,x}_\alpha \odot \vec{s}^{\,z}_\beta}$ is the bitwise binary dot product\,\footnote{The bitwise binary dot product is defined as ${\vec{s}^{\,x}_\alpha \odot \vec{s}^{\,z}_\beta} = \sum_i s^x_{i,\alpha}\odot s^z_{i,\beta}$ where $s^x_{i,\alpha}\odot s^z_{i,\beta}= \delta_{s^x_{i,\alpha},-1}\delta_{s^z_{i,\beta},-1}$.}. The IPR $\mathcal{I}^x$ is then given by
\eq{
  \mathcal{I}^x = \frac{1}{\nh^2}\sum_{\alpha}\big|\sum_\beta\psi_\beta^z \, (-1)^{\vec{s}^{\,x}_\alpha \odot \vec{s}^{\,z}_\beta}\big|^4\equiv \frac{1}{\nh^2}\sum_\alpha\Phi_\alpha\,.
  \label{eq:wh}
}
Provided the eigenstate is not fully delocalised in the $\sigma^z$-basis, then the only other way $\mathcal{I}^x\sim \nh^{-1}$ is that each $\Phi_\alpha$ in Eq.~\ref{eq:wh} has an $\mathcal{O}(1)$ contribution to the sum. This can happen only if $\psi^z_\beta$ is $\mathcal{O}(1)$ on only an $\mathcal{O}(1)$ number of $\vec{s}^{\,z}_\beta$ configurations; in other words, $\tau^z=0$. However, in the MBL-SG phase the eigenstates have fractal statistics in the $\sigma^z$-basis, $0<\tau^z<1$ strictly, which implies that $\tau^x<1$ also strictly.
Note that the discussions based on Eq.~\ref{eq:wht} and Eq.~\ref{eq:wh} are invariant under $x\leftrightarrow z$, which implies that $\tau^z<1$ strictly in the MBL-PM phase as well. This concludes the arguments behind the fractal structure of the states in either basis in the MBL-SG as well as in the MBL-PM phases.

In a nutshell, the analysis of the IPRs in both the bases show that their scaling with the Hilbert-space dimension can distinguish an MBL phase from an ergodic one, irrespective of in which basis the IPR is computed. However, it does not carry information of the specific phase structure in the MBL phase. In the following section, we will show how the spread of the eigenstates on the Hilbert space encodes precisely this information and how are they related to the order paramters characterising localisation protected order.

\section{Hilbert-space correlations and quantum order \label{sec:hscorrorder}}
In this section, we discuss the non-local correlations in the eigenstate amplitudes on the Hilbert space and show how they are encoded in measures of Hilbert-spatial spreads of eigenstates, which in turn are sensitive to order or lack thereof in the MBL phases.

Recall that $F^\mu(r)$ can be interpreted as a probability distribution over $r$ (see Eq.~\ref{eq:Fr}). The results in Fig.~\ref{fig:Fr} suggest that the mean position of this distribution as well as its variance would be revealing quantities to study. This is motivated by the observation that (i) $F^z(r^z)$ is unimodal in the MBL-PM as well as the ergodic phase whereas it is bimodal in the MBL-SG phase, and (ii) $F^x(r^x)$ is peaked at a $W_J/W_h$-dependent value in the MBL-PM phase but the peak gets pinned at $r^x=L/2$ throughout the ergodic and the MBL-SG phase.

In order to quantify these features, we define the mean position as 
\eq{
  \braket{r^\mu} = \sum_{r=0}^L r F^\mu(r)\,,
  \label{eq:r}
}
and the variance as
\eq{
  \braket{(\Delta^\mu_r)^2} = \sum_{r=0}^L r^2 F^\mu(r) - \bigg[\sum_{r=0}^L r F^\mu(r)\bigg]^2\,.
  \label{eq:rsq}
}

\subsection{Mean position $\braket{r^\mu}$}
\begin{figure}
\includegraphics[width=0.7\linewidth]{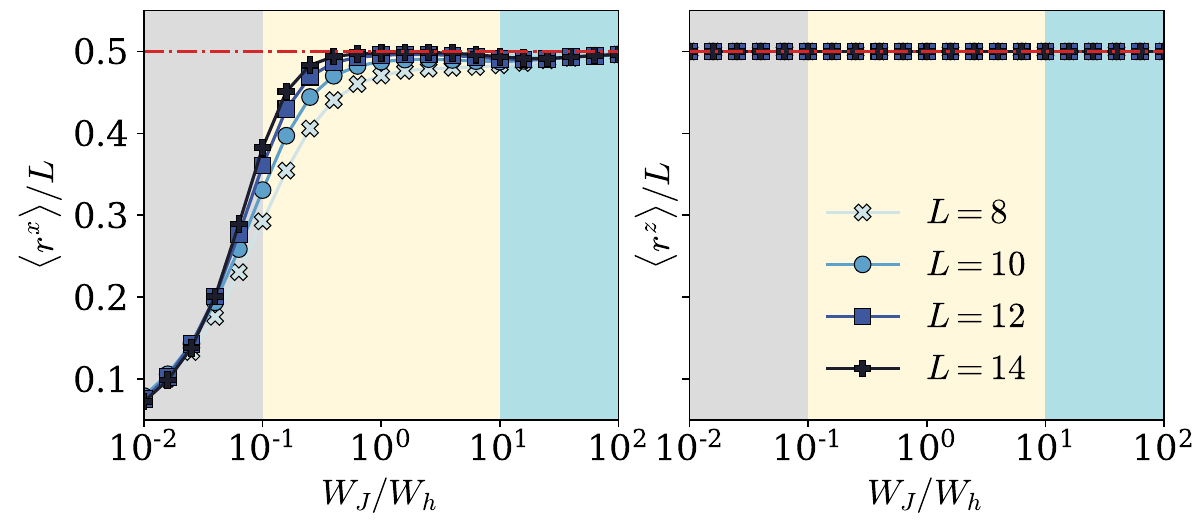}
\caption{The mean position, $\braket{r^\mu}/L$, defined in Eq.~\ref{eq:r}, as a function of $W_J/W_h$ for different system sizes $L$. The left panel corresponds to $\braket{r^x}/L$ whereas the right panel corresponds to $\braket{r^z}/L$.}
\label{fig:r}
\end{figure}

The results for $\braket{r^x}$ and $\braket{r^z}$ are shown in Fig.~\ref{fig:r} (left and right panels respectively). The ergodic phase is simple to understand as in this case, the eigenstates are very well approximated by Gaussian random vectors delocalised over the entire Hilbert-space graph, $|\psi^\mu_\alpha|^2|\psi^\mu_\beta|^2\sim \nh^{-2}$, 
such that $F^\mu(r)$ is dominated by $N_r=\binom{L}{r}$ which is sharply peaked at $r=L/2$, resulting in $\braket{r^\mu} = L/2$.

The MBL-PM and MBL-SG phases are more interesting. 
The MBL-PM eigenstates in the $\sigma^x$-basis are quite akin to conventional MBL eigenstates~\cite{roy2021fockspace}; the eigenstate correlations decay exponentially $|\psi^x_{\alpha}|^2|\psi^x_{\beta}|^2 \sim e^{-r^x_{\alpha\beta}/\xi^x}$ such that on an average we can write
\eq{
  F^x(r) = \mathfrak{N}^{-1}N_r e^{-r^x/\xi^x}\,,
}
where $\mathfrak{N}^{-1}$ is a normalisation constant. Recalling that $\sum_r F^\mu(r)=1$, the normalisation can be fixed as 
\eq{
  \mathfrak{N} = \sum_r N_r e^{-r^x/\xi^x} = (1+e^{-1/\xi^x})^L\,,
}
such that we have
\eq{
  F^x(r) = \binom{L}{r}\frac{e^{-r/\xi^x}}{(1+e^{-1/\xi^x})^L}\,,
  \label{eq:Fx-r-pm}
}
which implies
\eq{
  \braket{r^x}/L = (1+e^{1/\xi^x})^{-1}<1/2\,.  
}
In the MBL-SG phase on the other hand, $\braket{r^x}/L = 1/2$ even though the eigenstates have fractal statistics in the $\sigma^x$-basis. The physical picture that emerges out of this is that while the eigenstates are spatially spread out over the entire Hilbert-space graph, they are not spread out uniformly; in other words the wavefunction has holes in it, otherwise it spread all over the Hilbert space.

The situation with $\braket{r^z}$ is more subtle as $\braket{r^z}=L/2$ in the all the three phases but the anatomy of the eigenstaes is very different in the three phases. The behaviour in the ergodic phase is explained above. In the MBL-PM, eigenstates in the $\sigma^z$-basis are much like the MBL-SG eigenstates in the $\sigma^x$-basis; they are spread out over the entire Hilbert-space but inhomogeneously resulting in the weak multifractality, see Figs.~\ref{fig:Fr} and \ref{fig:ipr}. The anatomy is completely different in the MBL-SG phase. In the $\sigma^z$-basis the states are bilocalised on two configurations which are related to each other by the parity operator and hence distant by $r^z=L$. In the limit of $W_J\to\infty$, the eigenstates can be written as 
\eq{
  \ket{\psi} = \frac{1}{\sqrt{2}}[\ket{\vec{s}_\alpha^{\,z}}\pm P\ket{\vec{s}_\alpha^{\,z}}]\,,
}
where $r^z_{\alpha,\beta}=L$ with $\ket{\vec{s}_\beta^{\,z}}=P\ket{\vec{s}_\alpha^{\,z}}$. For such a state
\eq{
  F^z(r) = \frac{1}{2}(\delta_{r,0}+\delta_{r,L})\,,
}
such that $\braket{r^z}/L = 1/2$. For finite $W_J\gg W_h,W_V$, the eigenstates remain exponentially localised around $r=0$ and $r=L$ such that
\eq{
  F^z(r)=\mathfrak{N}^{-1}N_r[e^{-r/\xi^z}+e^{-(L-r)/\xi^z}]\,,
}
where the $\mathbb{Z}_2$ symmetry implies that both the peaks, around $r=0$ and $r=L$, have the same strength. The normalisation $\mathfrak{N}^{-1}$ can again be fixed by using $\sum_r F^z(r)=1$ which yields
\eq{
  F^z(r)=\frac{1}{2(1+e^{-1/\xi^z})^L}\binom{L}{r}[e^{-r/\xi^z}+e^{-(L-r)/\xi^z}]\,.
  \label{eq:Fz-r-sg}
}
The above form of $F^z(r)$ directly implies $\braket{r^z}/L = 1/2$ whereas the IPR, $\mathcal{I}^z = 2^{-1}(1+e^{-1/\xi^z})^{-L}$ which automatically implies the fractality.

Putting all the above together, we find that $\braket{r^x}/L = p<1/2$ in the MBL-PM phase whereas $\braket{r^x}/L=1/2$ otherwise, and $\braket{r^z}/L = 1/2$ in all three phases, see Fig.~\ref{fig:r}. In the above, we explained and understood these results based on the phenomenology of the eigenstates on the Hilbert space. In the following, we briefly discuss how this is microscopically related to the spin polarisations, along the lines of Ref.~\cite{roy2021fockspace}. The spin polarisation in an eigenstate $\ket{\psi}$ along the $\mu$-direction is defined as 
\eq{
  m^\mu &= L^{-1}\sum_{i=1}^L \braket{\psi|\sigma^\mu_i|\psi}^2\label{eq:mmudef}\\
  &= \sum_{\alpha,\beta}|\psi_\alpha^\mu|^2|\psi_\beta^\mu|^2 L^{-1}\sum_{i=1}^Ls^\mu_{i,\alpha}s^\mu_{i,\beta}\nonumber\\
  \Rightarrow m^\mu&= 1-\frac{2}{L}\braket{r^\mu}\label{eq:mmu-rmu}\,,
}
where in the last equality we used Eq.~\ref{eq:dist}. Note that the square of the expectation value in Eq.~\ref{eq:mmudef} makes $m^\mu$ insensitive to if the spin is polarised along the positive or negative $\mu$-direction. It thus makes $m^\mu$ a measure of how close the eigenstate is to a product state in the $\sigma^\mu$ direction which also, physically, is encoded in $\braket{r^\mu}$. 

That $\braket{r^x}/L = p<1/2$ in the MBL-PM phase implies, from Eq.~\ref{eq:mmu-rmu}, that the eigenstates have a finite polarisation along the $x$-direction, $m^x$, characteristic of a conventional MBL phase. In the MBL-SG phase, on the contrary, $\braket{r^x}=L/2$ implies that the eigenstates have no $\sigma^x$-polarisation, which is consistent with them being bilocalised in the $\sigma^z$-basis. In the same spirit, the MBL-PM eigenstates, which are polarised along the $\sigma^x$-axis, have no $\sigma^z$-polarisation consistent with $\braket{r^z}=L/2$. However, the mechanism by which the $\sigma^z$-polarisation vanishes in the MBL-SG phase is different; there since the eigenstates are approximate cat states of $\sigma^z$-configurations,
we have
\eq{
  \braket{\psi|\sigma^z_i|\psi}\approx \frac{s^z_{i,\alpha}}{2}(\bra{\vec{s}^{\,z}_\alpha}\pm\bra{\vec{s}^{\,z}_\alpha}P )(\ket{\vec{s}^{\,z}_\alpha}\mp P\ket{\vec{s}^{\,z}_\alpha} )=0\,,
}
which is again consistent with $\braket{r^z}=L/2$.

To summarise this section, we find that the mean position $\braket{r^\mu}$ directly encodes the eigenstate's spin polarisation $m^\mu$. However, as we explained, this measure is rather insufficient to detect localisation protected order as $\braket{r^z}/L=1/2$ in all three phases whereas $\braket{r^x}/L=p<1/2$ only in the MBL-PM phase ($1/2$ otherwise), which is a feature of conventional MBL phases without any spontaneously symmetry broken order.

In the next section, we show that the variances, $\braket{(\Delta^\mu_r)^2}$, defined in Eq.~\ref{eq:rsq}, in fact carry the signatures of the quantum order and do indeed distinguish between the two localised phases.

\subsection{Variances $\braket{({\Delta}^\mu_r)^2}$}

\begin{figure}[!t]
\includegraphics[width=0.7\linewidth]{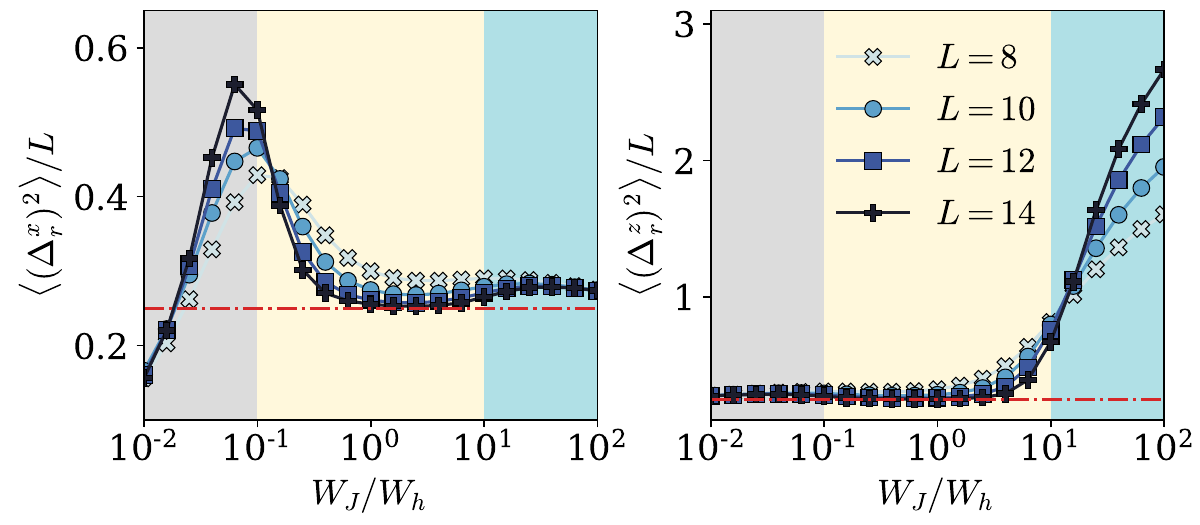}
\caption{The variances, $\braket{(\Delta_r^\mu)^2}/L$, defined in Eq.~\ref{eq:rsq}, as a function of $W_J/W_h$ for different system sizes $L$. The left panel corresponds to $\braket{(\Delta_r^x)^2}/L$ whereas the right panel corresponds to  $\braket{(\Delta_r^z)^2}/L$. The red dashed line corresponds to the value of 1/4.}
\label{fig:rsq}
\end{figure}

The results for the variances, $\braket{({\Delta}^\mu_r)^2}$, are shown in Fig.~\ref{fig:rsq}. Within the MBL-PM phase, in the $\sigma^x$-basis, the form of $F^x(r)$ in Eq.~\ref{eq:Fx-r-pm} implies that 
\eq{
  \braket{(\Delta^x_r)^2}/L = e^{1/\xi^x}(1+e^{1/\xi^x})^{-2}\,,
}
which formally explains why $\braket{(\Delta^x_r)^2}$ scales with $L$ inside the phase. In the ergodic and MBL-SG phases, the eigenstates are spread out over the entire Hilbert-space graph in the $\sigma^x$-basis, albeit with holes in the MBL-SG phase. Formally, this is equivalent to setting $\xi^x\to\infty$ which implies $\braket{(\Delta^x_r)^2}=L/4$ in these two phases. This is indeed confirmed by the numerical results in the left panel in Eq.~\ref{fig:rsq}. Note that all three phases, $\sqrt{\braket{(\Delta^x_r)^2}}\propto \sqrt{L}$ whereas the mean position $\braket{r^x}\propto L$ as well; this means in the $\sigma^x$-basis the $F^x(r)$ remains infinitely sharply peaked around its mean position in the thermodynamic limit. The only exception to this is at the MBL-PM to ergodic phase transition where $\braket{(\Delta^x_r)^2}/L$ seems to grow with $L$; we shall shortly explain this in terms of a divergent real-space correlation length.

Turning to the $\sigma^z$-basis, in the MBL-SG phase, the form of $F^z(r)$ in Eq.~\ref{eq:Fz-r-sg} implies that
\eq{
  \braket{(\Delta^z_r)^2}/L = \frac{L}{4}\frac{(1-e^{1/\xi^z})^2}{(1+e^{1/\xi^z})^2} +\frac{e^{1/\xi^z}}{(1+e^{1/\xi^z})^2}\,,
  \label{eq:var-z}
}
which is extremely important as the $\sim L^2$ scaling of $\braket{(\Delta^z_r)^2}$ hitherto the first clear signature of the MBL-SG phase in Hilbert-space correlations. Physically, this is a direct manifestation of the eigenstates being bilocalised on two $\sigma^z$-configurations which are separated by distance $r=L$. As such, the variance in the distribution $F^z(r)$ naturally scales as $L^2$. Again, in the ergodic and MBL-PM phases, the eigenstates are delocalised over the entire Hilbert-space, broadly speaking, such that one can nominally set $\xi^z\to\infty$ in which case one obtains $\braket{(\Delta^z_r)^2} = L/4$. These results are indeed corroborated by the numerical results in the right panel in Fig.~\ref{fig:rsq}.

The central result in this section is that in the MBL-SG phase $\braket{(\Delta^z_r)^2}\sim L^2$, which is exclusive to the MBL-SG phase as in every other phase both $\braket{(\Delta^z_r)^2}$ and $\braket{(\Delta^x_r)^2}$ scale as $L$. Underpinning this is a microscopic connection between the variance and the Edwards-Anderson order parameter defined in Eq.~\ref{eq:EA}, with the latter in turn diagnosing the MBL-SG phase.

In order to derive this connection, note that from Eq.~\ref{eq:dist} we have
\eq{
  (r^\mu_{\alpha\beta})^2 = \frac{L^2}{4}+\frac{1}{4}\sum_{i,j}s^\mu_{i,\alpha}s^\mu_{j,\alpha}s^\mu_{i,\beta}s^\mu_{j,\beta}-\frac{L}{2}\sum_{i}s_{i,\alpha}^\mu s_{i,\beta}^\mu\,,
}
such that
\eq{
  \braket{(r^\mu)^2}=\frac{L^2}{4}+\frac{1}{4}\sum_{i,j}\braket{\sigma^\mu_i\sigma^\mu_j}^2-\frac{1}{2}\sum_{i}\braket{\sigma^\mu_i}^2\,.
}
Using the above and Eq.~\ref{eq:mmu-rmu} in Eq.~\ref{eq:rsq}, we have
\eq{
\frac{\braket{(\Delta^\mu_r)^2}}{L}=\frac{1}{4L}\bigg[\sum_{i,j}\braket{\sigma^\mu_i\sigma^\mu_j}^2-\sum_{i,j}\braket{\sigma^\mu_i}^2\braket{\sigma^\mu_j}^2\bigg]\,.
\label{eq:rsd-corr}
}
In all the three phases we have $\braket{\sigma^z_i} = 0$ albeit for different reasons. In the MBL-PM phase, it is so as the states are approximately polarised in the $\sigma^x$-direction and hence are equal superpostions of the $\sigma^z=1$ and $-1$ states. In the ergodic phase, ETH mandates that $\braket{\sigma^z_i} = 0$ as we are at infinite temperatures. Finally, in the MBL-SG phase $\braket{\sigma^z_i} = 0$ is effected by the broken $\mathbb{Z}_2$ symmetry.
This leads to the simplification that 
\eq{
  \frac{\braket{(\Delta^\mu_r)^2}}{L}=\frac{1}{4L}\sum_{i,j}\braket{\sigma^\mu_i\sigma^\mu_j}^2 = \chi/4\,,
  \label{eq:var-chi}
}
where the last equality follows from the definition in Eq.~\ref{eq:EA}.
Since the Edwards-Anderson order parameter $\chi\propto L$ in the MBL-SG phase, we have $\braket{(\Delta^z_r)^2}\propto L^2$. This is due to the fact that there is long-range spin-glass order such that $\braket{\sigma^z_i\sigma^z_j}^2\sim \mathcal{O}(1)$ even if $|i-j|\to \infty$ in a thermodynamically large system. On the other hand, in the absence of long-range order, we expect $\braket{\sigma^z_i\sigma^z_j}^2\sim e^{-|i-j|/\zeta^z}$, such that $\chi\sim \mathcal{O}(1)$ and $\braket{(\Delta^z_r)^2}\sim L$, as is the case in the ergodic and MBL-PM phases, see Fig.~\ref{fig:rsq}. The relation between the variance and the Edwards-Anderson paramter in Eq.~\ref{eq:var-chi} along with the expression for the variance in Eq.~\ref{eq:var-z} yields a relation for the Edwards-Anderson order parameter in terms of a correlation length, $\xi^z$, on the Hilbert space. It is consistent that as this correlation length diverges, which one expects at the MBL-SG to ergodic phase transition, the coefficient of the term in $\chi$ scaling as $L^2$ vanishes.

Note that, from Eq.~\ref{eq:rsd-corr}, the variance can be expressed as a correlation function $\braket{(\Delta^\mu_r)^2}/L = \sum_{i,j}C_{ij}/4L$, where $C_{ij}=\braket{\sigma^\mu_i\sigma^\mu_j}^2-\braket{\sigma^\mu_i}^2\braket{\sigma^\mu_j}^2$ can be interpreted as a correlation between polarisations at sites $i$ and $j$. For short-range correlations, $C_{ij}\sim e^{-|i-j|/\zeta^\mu}$ we have $\braket{(\Delta^\mu_r)^2}\sim L$ whereas if the correlations become long-ranged, $\braket{(\Delta^\mu_r)^2}\sim L^{\alpha>1}$ which is what happens at with $\mu=x$ at the MBL-PM to ergodic transition. This suggests that the correlation length associated with $C_{ij}$ with $\mu=x$ diverges at the transition.

\begin{table}[!t]
\begin{tabular}{c||c|c|c}
 & MBL-PM & Ergodic & MBL-SG \\
\hline
\hline
$\langle r^x \rangle/L=$ & $p<1/2$& 1/2& 1/2\\
$\langle r^z \rangle/L=$ &1/2 &1/2 &1/2 \\
$\langle (\Delta_r^x)^2 \rangle/L=$ & $\mathcal{O}(1)$&1/4 &$\approx 1/4$ \\
$\langle (\Delta_r^z)^2 \rangle/L=$ &$\approx 1/4$ &1/4 & $\mathcal{O}(L)$\\
\end{tabular}
\caption{Table summarising the mean positions $\braket{r^\mu}/L$ and the variances $\braket{(\Delta_r^\mu)}^2/L$ in the three phases.}
\label{tab:r-rsq}
\end{table}

We close this section with summarising the mean positions $\braket{r^\mu}/L$ and the variances $\braket{(\Delta_r^\mu)}^2/L$ in the three phases in Table~\ref{tab:r-rsq}.

\section{Hilbert-space correlations and entanglement \label{sec:hscorrent}}
The results above showed that appropriate Hilbert-space correlations not only distinguish different MBL phases from the ergodic phase, but can also characterise the presence or absence of localisation protected quantum order in the MBL phases and hence distinguish ordered MBL phases from those without order. However, another important aspect of MBL phases that distinguish them from ergodic ones are their entanglement structure. MBL eigenstates are characterised by area-law entanglement~\cite{serbyn2013local,bauer2013area,luitz2015many,geraedts2016many,geraedts2017characterising} as opposed to the default volume-law for ergodic eigenstates. In a recent work~\cite{roy2022hilbert}, it was shown that appropriately defined four-point Hilbert-space correlations of eigenstate amplitudes encode the information of bipartite entanglement. It is, therefore, worth asking how these four-point correlations, in both the $\sigma^z$- and $\sigma^x$-bases, behave across the two transitions in the model in Eq.~\ref{eq:ham}.

While the details can be found in Ref.~\cite{roy2022hilbert}, we briefly recapitulate the connection between the four-point correlation and a bipartite-entanglement measure here for completeness. Instead of the conventionally studied, von Neumann entropy of entanglement, we consider the bipartite purity between subsystem $A$ and its complement $B$. For simplicity, we split the chain into two equal halves and denote the left half as $A$ with length $L_A=L/2$ and the right half as $B$ with $L_B=L/2$. For a state $\rho=\ket{\psi}\bra{\psi}$, it is defined as 
\eq{
  \mathcal{P} = \mathrm{Tr}_A[(\mathrm{Tr}_B \rho)^2]\,,
}
where $\mathrm{Tr}_B$ denotes the partial trace over subsystem $B$ and $\rho_A\equiv\mathrm{Tr}_B\rho$ is the reduced density matrix of subsystem $A$. For volume-law entangled states, $\rho_A$ is heavily mixed, such that $\mathcal{P}$ is exponentially small in $L$. On the other hand, for area-law entangled states, $\rho_A$ retains its purity such that  $\mathcal{P}\sim \mathcal{O}(1)$ in the thermodynamic limit.

Schmidt-decomposing a state $\ket{\psi}$ in the $\sigma^\mu$-basis as 
\eq{
  \ket{\psi} = \sum_{\alpha_A,\alpha_B}\psi_{\alpha_A\alpha_B}^\mu\ket{\vec{s}^{\,\mu}_{\alpha_A}}\otimes\ket{\vec{s}^{\,\mu}_{\alpha_B}}\,,
  \label{eq:schmidt}
}
where $\ket{\vec{s}^{\,\mu}_{\alpha_{A(B)}}}$ denotes the spin-configuration in subsystem $A(B)$, the purity can be written as 
\eq{
  \mathcal{P} = \sum_{r_A=0}^{L_A}\sum_{r_B=0}^{L_B}F^\mu_\mathcal{P}(r_A,r_B)\,,
  \label{eq:P-Frarb}
}
with the two-distance Hilbert-spatial correlation
\eq{
  F^\mu_\mathcal{P}(r_A,r_B) = \sum_{\substack{\alpha_A,\beta_A: \\r^\mu_{\alpha_A\beta_A}=r_A}}\sum_{\substack{\alpha_B,\beta_B: \\r^\mu_{\alpha_B\beta_B}=r_B}}& \bigg[\psi_{\alpha_A\alpha_B}^\mu (\psi_{\beta_A\alpha_B}^\mu)^\ast \times\nonumber\\&(\psi_{\alpha_A\beta_B}^\mu)^\ast \psi_{\beta_A\beta_B}^\mu\bigg]\,.
  \label{eq:Frarb}
}
In the above, $r^\mu_{\alpha_A\beta_A}$ denotes the Hamming distance in subsystem $A$ and similarly for $r^\mu_{\alpha_B\beta_B}$.

In Ref.~\cite{roy2022hilbert} it was shown that $\mathcal{P}^{-1}F^\mu_\mathcal{P}(r_A,r_B)$ could be thought of as a distribution over the $(r_A,r_B)$ space, and this distribution had qualitatively different behaviour in an ergodic and an MBL phase. In the former, it is supported only on $r_A=0$ and $r_B=0$. As such, it has a form
\eq{
  F^\mu_\mathcal{P}(r_A,r_B) = \nh^{-1}\left[\delta_{r_A0}\delta_{r_B0} + \binom{L_A}{r_A}\delta_{r_B0}+\binom{L_B}{r_B}\delta_{r_A0}\right]\,,
  \label{eq:Frarb-erg}
}
whereas in the latter, it is supported on an extensive $(r_A,r_B)$ and given by
\eq{
  F^\mu_\mathcal{P}(r_A,r_B) = \binom{L_A}{r_A}\binom{L_B}{r_B}s^{r_A+r_B}(1-s)^{L-r_A-r_B}\,,
  \label{eq:Frarb-MBL}
}
where $s$ decays towards 0 as the system is driven deeper into the MBL phase. The form in Eq.~\ref{eq:Frarb-MBL} also implies that $F^\mu_\mathcal{P}(r_A,r_B)$ is sharply peaked around $(r_A,r_B)=(sL_A,sL_B)$ in the MBL phase. The two behaviours can be quantitatively distinguished by defining a quantity
\eq{
  R^\mu &= \frac{\braket{r_Ar_B}}{\braket{r_A+r_B}^2}\nonumber\\& \equiv \frac{\mathcal{P}^{-1}\sum_{r_A,r_B}r_Ar_BF^\mu_\mathcal{P}(r_A,r_B)}{\left[\mathcal{P}^{-1}\sum_{r_A,r_B}(r_A+r_B)F^\mu_\mathcal{P}(r_A,r_B)\right]^2}\,,
  \label{eq:R-def}
}
which vanishes as $L\to\infty$ in the ergodic phase and saturates to an $\mathcal{O}(1)$ number in the MBL phase. In particular, $R^\mu$ saturates to $1/4$ in the infinite disorder limit. This concludes our recapitulation of the characterisation of entanglement structure from Hilbert-space correlations in conventional MBL systems. We next discuss our results for the same for the model in Eq.~\ref{eq:ham} which shows localisation protected order.

\begin{figure}
\includegraphics[width=0.7\linewidth]{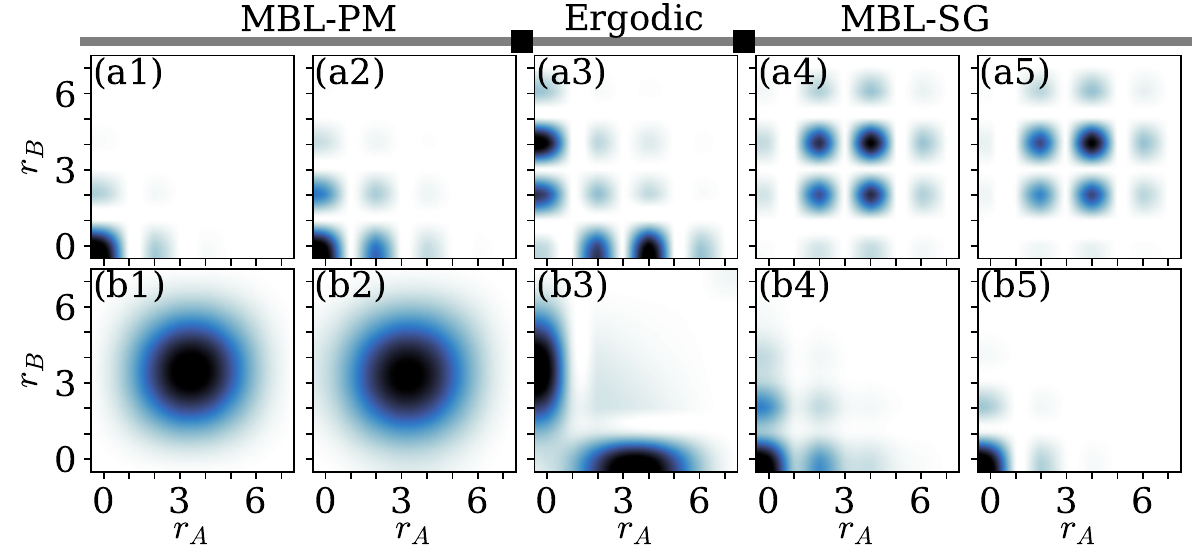}
\caption{The two-distance Hilbert-spatial correlation function, $F^\mu_\mathcal{P}$, defined in Eq.~\ref{eq:Frarb} for several parameter values spanning the three phases of the model in Eq.~\ref{eq:ham} shown as heatmaps with darker colours denoting higher values and white denoting zero. The top and the bottom rows correspond to $F^x_\mathcal{P}$ and $F^z_\mathcal{P}$ respectively. The parameter across the five columns are $W_J/W_h = 0.01$, $0.06$, $1.0$, $16.0$, and $100$ respectively. The leftmost two columns correspond to the MBL-PM phase, the rightmost two columns to the MBL-SG phase, and the central column to the ergodic phase. Data shown is for $L=14$.}
\label{fig:Frarb}
\end{figure}

A broadbrush view of $F^\mu_\mathcal{P}(r_A,r_B)$ as heatmaps in the $(r_A,r_B)$ plane for different parameter values spanning the three phases is shown in Fig.~\ref{fig:Frarb}. The top row shows $F^x_\mathcal{P}$ and the bottom row shows $F^z_\mathcal{P}$. The data in panels (a3) and (b3), corresponding to the ergodic phase, are entirely consistent with Eq.~\ref{eq:Frarb-erg}. Similarly, $F^x_\mathcal{P}(r_A,r_B)$ in the MBL-PM phase [panels (a1) and (a2)] and $F^z_\mathcal{P}(r_A,r_B)$ in the MBL-SG phase (panels (b4) and (b5)) are consistent with the form in Eq.~\ref{eq:Frarb-MBL}. Note that the vanishing $F^\mu_\mathcal{P}(r_A,r_B)$ at odd values of $r_A$ or $r_B$ are simply a consequence of the eigenstates also being parity eigenstates. These results, so far, are quite similar in spirit to those in conventional MBL systems~\cite{roy2022hilbert}. 

However, what is more intriguing and novel is the behaviour of $F^x_\mathcal{P}(r_A,r_B)$ in the MBL-SG phase [panels (a4) and (a5)] and that of $F^z_\mathcal{P}(r_A,r_B)$ in the MBL-PM phase [panels (b1) and (b2)]. Despite the fact that the MBL-SG and MBL-PM eigenstates are quite delocalised over the Hilbert-space in the $\sigma^x$- and $\sigma^z$-bases respectively, as evinced by $\braket{r^x}=L/2$ and $\braket{r^z}=L/2$ in the two cases, the corresponding $F^\mu_\mathcal{P}(r_A,r_B)$ are supported on $r_A\approx L_A/2$ and $r_B\approx L_B/2$. This indicates a high degree of correlation among the eigenstate amplitudes. One way to understand this is that, in the ergodic phase, the fact that the eigenstates can be approximated as Gaussian random vectors with each element independently distributed, as opposed to being correlated, is what forces either $r_A=0$ or $r_B=0$ in Eq.~\ref{eq:Frarb} leading to the result in Eq.~\ref{eq:Frarb-erg}. Only strong correlations in the eigenstate amplitudes can lead to the summands with $r_A,r_B\neq0$ having a finite contribution to the sum in Eq.~\ref{eq:Frarb}.

\begin{figure}[!b]
\includegraphics[width=0.7\linewidth]{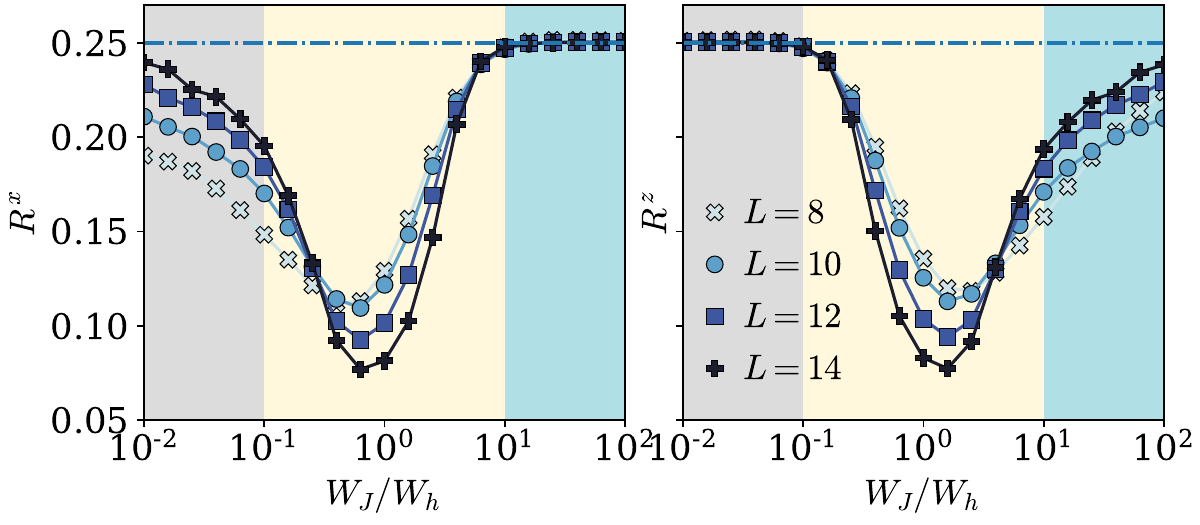}
\caption{$R^x$ (left) and $R^z$ (right) defined in Eq.~\ref{eq:R-def} as a function of $W_J/W_h$ for the different system sizes $L$. In the ergodic phase, both of them decay towards zero with increasing $L$ whereas in MBL phases, they seemingly saturate to a finite constant with increasing $L$. The blue dashed lines denote the value of 1/4 which is expected deep in either MBL phase.}
\label{fig:R-ent}
\end{figure}

The above arguments can be made more explicit by considering a couple of toy examples. As a limiting case of an MBL-PM eigenstate, consider $\ket{\psi} = \ket{\vec{s}^{\,x}_\alpha}$ which is a product state in the $\sigma^x$-basis that satisfies $\sigma^x_i\ket{\psi}=\ket{\psi}$. For this state $F^x_\mathcal{P}(r_A,r_B)=\delta_{r_A0}\delta_{r_B0}$. However, the situation is richer in the $\sigma^z$-basis. Using Eq.~\ref{eq:wht}, the amplitudes in the $\sigma^z$-basis can be written as
\eq{\psi_\beta^z = \nh^{-1/2}(-1)^{\vec{s}^{\,x}_\alpha \odot \vec{s}^{\,z}_\beta}=\nh^{-1/2}\,,}
Using this in Eq.~\ref{eq:Frarb}, we immediately have
\eq{
  F^z_\mathcal{P}(r_A,r_B) = \nh^{-1}\binom{L_A}{r_A}\binom{L_B}{r_B}\,,
}
which is entirely in consonance with the data in panels (b1) and (b2) in Fig.~\ref{fig:Frarb}.

Similarly, a cartoon MBL-SG eigenstate can be written as 
\eq{
  \ket{\psi} = \frac{1}{\sqrt{2}}[\ket{\vec{s}_\alpha^{\,z}}+ P \ket{\vec{s}_\alpha^{\,z}}]\,,
}
where $\sigma^z_i\ket{\vec{s}_\alpha^{\,z}}=\ket{\vec{s}_\alpha^{\,z}}$
Again using Eq.~\ref{eq:wht}, the amplitudes in the $\sigma^x$-basis can be expressed as 
\eq{
  \psi_\beta^x = \frac{1}{\sqrt{2\nh}}[(-1)^{\vec{s}^{\,x}_\beta \odot \vec{s}^{\,z}_\alpha} + (-1)^{\vec{s}^{\,x}_\beta \odot \overline{\vec{s}^{\,z}_\alpha}}]\,,
}
where $\overline{\vec{s}^{\,z}_\alpha}$ denotes the configuration with all spins flipped relative to $\vec{s}^{\,z}_\alpha$. The above form immediately implies that 
\eq{
  \psi_\beta^x=\begin{cases}
  0; & n_{\beta,-1} = \mathrm{odd}\\
  \sqrt{2/\nh}; & n_{\beta,-1} = \mathrm{even}
  \end{cases}\,,
  \label{eq:sg-cartoon}
}
where $n_{\beta,-1}$ is the number of real-space sites where $\sigma^x_i\ket{\vec{s}^{\,x}_\beta} = -\ket{\vec{s}^{\,x}_\beta}$. Note that the amplitudes in Eq.~\ref{eq:sg-cartoon} ensure that the eigenstate is indeed a parity eigenstate. It also implies that the state is uniformly spread over all $\sigma^x$-configurations which reside in a fixed parity sector. Since, all such configurations have an even Hamming distance $r^x$ we have, using Eq.~\ref{eq:sg-cartoon} in Eq.~\ref{eq:Frarb}
\eq{
  F^x_\mathcal{P}(r_A,r_B) = \begin{cases}\frac{4}{\nh}\binom{L_A}{r_A}\binom{L_B}{r_B}; & r_A,r_B = \mathrm{even}\\
  0; &\mathrm{otherwise}
  \end{cases}\,,
}
which is again consistent with the data in in panels (a4) and (a5) in Fig.~\ref{fig:Frarb}.

To summarise, the two-distance Hilbert-spatial correlation, $F^\mu_\mathcal{P}(r_A,r_B)$, in Eq.~\ref{eq:Frarb} is finite only for $r_A=0$ or $r_B=0$ in the ergodic phase for both the basis choices. However, in both the MBL-PM and MBL-SG phases, the correlation is supported on finite and extensive $r_A,r_B$ in either basis. This suggests that quantifying this difference via the quantity defined in Eq.~\ref{eq:R-def} will be useful. We show $R^x$ and $R^z$ in the left and right panels, respectively, of Fig.~\ref{fig:R-ent}. In the ergodic phase, indeed, $R^\mu$ decays towards zero with increasing $L$, whereas in the MBL phases, it seems to saturate to a finite value with increasing $L$ with saturation value tending towards 1/4 as one goes deeper into the MBL phases. This concludes our discussion of the connection between Hilbert-space correlations and entanglement structures across the two phase transitions in the model in Eq.~\ref{eq:ham}.

\section{Conclusions \label{sec:conclusions}}
To summarise, we analysed the Hilbert-space anatomy of eigenstates in MBL phases with and without localisation protected order as well in the intervening ergodic phase in a $\mathbb{Z}_2$ symmetric, disordered Ising spin chain, described by the Hamiltonian in Eq.~\ref{eq:ham}. We demonstrated that the spread of the eigenstates on the Hilbert-space graph quantified in terms of non-local correlations of eigenstate amplitudes, carried the information of whether the system was in an MBL phase with or without spin-glass order or in an ergodic phase. In particular, we defined a spatial correlation on the Hilbert-space graph, $F^\mu(r)$, (see Eq.~\ref{eq:Fr}) where the Hamming distance $r$ endowed the graph with a natural notion of distance. We found that in the MBL-PM phase the eigenstates are conventionally localised in the $\sigma^x$-basis resulting in a single peak in the correlations near $r=0$. On the other hand, in the MBL-SG phase, the `feline' nature of the eigenstates in the $\sigma^z$-basis manifested itself in a bilocalised structure such that the correlations are peaked near $r=0$ and $r=L$.
The MBL-PM and MBL-SG eigenstates in the $\sigma^z$- and $\sigma^x$-basis respectively, are spatially spread out over the Hilbert space but with `holes'. This results in a weak multifractality of the eigenstates which could be understood 
by virtue of them being related to their strongly multifractal counterparts in the $\sigma^x$- and $\sigma^z$-basis respectively via Walsh-Hadamard transforms.

We argued that $F^\mu(r)$ could be interpreted as a normalised distribution over all $r$ such that one could define a mean position $\braket{r^\mu}$ and a variance $\braket{(\Delta^\mu_r)^2}$ (see Eq.~\ref{eq:r} and Eq.~\ref{eq:rsq}). 
The main result was that $\braket{r^\mu}$ was directly related to the $\sigma^\mu$-spin polarisations in the eigenstates, and $\braket{(\Delta^\mu_r)^2}$ directly encodes the Edwards-Anderson spin-glass order parameter, again for the $\sigma^\mu$-operators. These measures of the Hilbert-spatial position and spread of the eigenstates thus characterise the three phases. In particular, we showed that $\braket{(\Delta^z_r)^2}=\chi L/4$ with $\chi$ the Edwards-Anderson order parameter. The bilocalised structure of the MBL-SG eigenstates on states separated by $r^z=L$ implies that $\braket{(\Delta^z_r)^2}\sim L^2$ which in turn means that $\chi\sim L$ indicating the MBL-SG phase.
We also characterised the entanglement structure in the three phases, area-law in the two MBL phases and volume-law in the ergodic phase, in terms of higher-point Hilbert-space correlations, $F^\mu_\mathcal{P}(r_A,r_B)$ (see Eq.~\ref{eq:Frarb}). We find that the correlations $F^\mu_\mathcal{P}(r_A,r_B)$ in both the bases carry the information of the entanglement transitions.

A question of immediate future interest is to analyse the critical properties of the two transitions in terms of scaling of the Hilbert-space lengthscales. In Refs.~\cite{mace2019multifractal,roy2021fockspace}, such an approach lead to a Kosterlitz-Thouless-like scenario for the `conventional' many-body localisation transition which was accompanied by a discontinuous jump of the spin-polarisations. The framework in this work can be used to develop an analogous picture for the $\mathbb{Z}_2$ symmetric case and gain insights into the behaviour of the Edwards-Anderson spin-glass parameter across the MBL-SG to ergodic phase transition.
Furthermore, since these phases and the phase transitions are fundamentally dynamical in nature, their dynamical signatures but from the Hilbert-space viewpoint is another interesting question. In particular, starting from some simple state, one can ask how does the wavefunction spread on the Hilbert-space graph~\cite{creed2023probability} and how does that carry the information of localisation protected order or lack thereof.

Finally, in this work, we considered possibly the simplest case of $\mathbb{Z}_2$ symmetry. An important question that arises is regarding the fate of many-body localisation in systems with non-Abelian symmetries. In particular, it is understood that MBL phases are in fact incompatible with non-Abelian symmetries~\cite{potter2016symmetry} and an MBL phase in such systems is necessarily accompanied by a spontaneous breaking of the symmetry~\cite{prakash2017eigenstate,friedman2018localisation}. Understanding this phenomenon and its manifestation on the Hilbert-space anatomy of eigenstates is a question of immanent interest.

\section*{Acknowledgements}
SR acknowledges support from the Department of Atomic Energy, Government of India, under project no. RTI4001 and from an ICTS-Simons Early Career Faculty Fellowship via a grant from the Simons Foundation (677895, R.G.). All numerical computations were performed on the \texttt{NEMO} workstation at ICTS-TIFR.

\bibliography{temp}

\end{document}